\documentclass[]{aa}
\usepackage{psfig}

\begin{document}

\thesaurus {11.05.2, 11.07.1, 11.09.2, 11.16.1, 11.16.2, 11.19.2}

\title{ Statistics of galaxy warps in the HDF North and South}
\titlerunning{Warps}

\author{V. Reshetnikov\inst{1,5}, E. Battaner\inst{2},
F. Combes\inst{3}, \and  J.  Jim\'{e}nez-Vicente\inst{4} }

\offprints{V.P. Reshetnikov~resh@astro.spbu.ru }

\institute{Astronomical Institute of St.Petersburg State
University, 198504 St.Petersburg, Russia
\and Depto. F\'{i}sica Te\'{o}rica y del Cosmos, Universidad de
Granada, Spain
\and DEMIRM, Observatoire de Paris, 61 Av. de l'Observatoire, 
75014 Paris, France
\and Groningen Kapteyn Laboratorium, Groningen, Netherlands
\and Isaac Newton Institute of Chile, St.Petersburg Branch}

\date{Received 6 September 2001; Accepted 13 November 2001}

\maketitle

\abstract{
We present a statistical study of the presence of galaxy warps in the 
Hubble deep fields. Among a complete sample of 45 edge-on galaxies
above a diameter of 1.$''$3, we find 5 galaxies to be certainly warped
and 6 galaxies as good candidates. In addition, 4 galaxies reveal
a characteristic U-warp. Compared to statistical studies of local
warps, and taking into account the strong bias against
observing the outer parts of galaxies at high redshift,
these numbers point towards a very high frequency of warps at
$z \sim 1$: almost all galaxy discs might be warped.
Furthermore, the amplitude of warps are stronger than for
local warps. This is easily interpreted in terms of higher 
galaxy interactions and matter accretion in the past.
This result supports these two mechanisms as the best 
candidates for the origin of early warps.  
The mean observed axis ratio of our sample of edge-on galaxies
is significantly larger in the high-z sample than is found for
samples of local spiral galaxies.
This might be due to disk thickening
due to more frequent galaxy interactions.
}

\section{Introduction}

Warping of outer galactic disks is a very common phenomenon,
with amplitude increasing with radius. Warps are
mostly conspicuous in the HI plane, since the atomic gas extends
up to 100kpc, and its radius is often several times that of the optical disk
(e.g. Bosma 1978, Briggs 1990). However warps start at the outskirts
of optical disks, and careful investigation of edge-on galaxies have
revealed that a significant percentage (40-50\%) of stellar disks are also
warped (S\'{a}nchez-Saavedra et al. 1990, Reshetnikov \& Combes 1998).

The origin of warps has been a puzzle for a long time, since orbits
slightly inclined with respect to the plane precess around the disk 
rotation axis,
and the precession is differential: the life-time of warps would
be much less than the Hubble time. Several mechanisms to trigger and
maintain warps have been proposed, including tidal interactions between
galaxies, inter-galactic magnetic fields, disk-halo interaction and matter
accretion (see the review by Binney, 1992). The forcing of a misaligned halo
does not appear to be a promising solution, since the halo and disks align
over a short time-scale (Nelson \& Tremaine 1995, Binney et al. 1998).
On the contrary, galaxy interactions have been revived as an explanation,
since even the prototypical isolated warped galaxy NGC 5907 has been
found to interact with small companions (Shang et al. 1998).
The other likely alternative is gas accretion, since it has been realized that
galaxies are still significantly growing in mass during all their lives, and
are surrounded by a network of filaments accounting for the Lyman-$\alpha$ absorbers
(e.g. Dav\'e et al. 1999). The outer parts of galaxies could be constantly
accreting gas at an inclined angular momentum, explaining the persistence of
warps. The dark matter is also constantly raining down onto the
dark galactic halo, with a skewed angular momemtum, making the galaxy
re-orient completely in a few Gyrs (Jiang \& Binney 1999).

If these two mechanisms, matter accretion and tidal interaction with
companions, are the main drivers of warps, it is very likely that
warps were even stronger and more frequent in the past. 
Knowing this frequency at high redshift would help to 
discriminate between models of the warp origin.  
We thus undertake such an investigation in the Hubble Deep
Fields.
The main goal of our work is to study the frequency of optical
warps at $z \sim 1$. At present, HI-21cm observations are not
sensitive enough (and have insufficient spatial resolution) to 
probe high redshift galaxies.

Throughout this paper, we adopt a flat
cosmology with $q_0=0.5$ and $H_0=70$ km\,s$^{-1}$\,Mpc$^{-1}$.
If not differently specified, magnitudes are expressed in the AB
system (Oke 1974).

\section{The sample}

The Hubble deep fields north and south (HDF-N and HDF-S, respectively)
are the best available sources of data on the distant
Universe (see Ferguson et al. 2000 for a comprehensive review).
The two fields give the possibility to study distant field galaxies 
to the faintest levels currently achievable. Therefore, we
selected HDF-N and HDF-S to study optical warps in remote galaxies.

\subsection{Selection}

We select our sample of distant edge-on galaxies on the basis
of visual inspection of original HDF-N and HDF-S frames in the F814W 
filter (hereafter $I_{814}$). Visual selection is probably the 
best way to recognize true strongly inclined spirals among a huge number
of asymmetric and peculiar objects in the two crowded deep fields.

The resulting sample of selected objects consisted of 50 edge-on
galaxies. Tables 1 and 2 summarize the main characteristics
of the galaxies.

The columns of Table 1 are: number according to the catalogue of 
Fern\'{a}ndez-Soto et al. (1999); 
coordinates (in pixels); Williams et al. (1996) identification; $I_{814}$ 
magnitude in the $AB$ photometric system; spectroscopic
redshift (Cohen et al. 2000; Hogg et al. 2000); 
photometric redshift (Fern\'{a}ndez-Soto et al. 1999); photometric redshift 
according to Fontana et al. (2000); best-fit 
spectral type of galaxy (Fern\'{a}ndez-Soto et al. 1999); index of warp
(see sect.3.2).

The columns of Table 2 are: the name according to the Stony Brook catalogue 
(Fern\'{a}ndez-Soto al. 1999) (the numbers mean coordinates in pixels); 
$I_{814}$ magnitude in the $AB$ photometric system; 
photometric redshift (Fern\'{a}ndez-Soto et al. 1999); photometric redshift 
according to Fontana et al. (2000); best-fit 
spectral type of galaxy (Fern\'{a}ndez-Soto et al., 1999); index of warp.

\begin{table*}
\caption{Edge-on galaxies in the HDF-N}
\begin{center}
\begin{tabular}{|c|c|c|c|c|c|c|c|c|}
\hline
$N$ & Mosaic (X,Y)&Williams et al.& $I_{814}$& $z$ & $z_{fsoto}$ & 
$z_{font}$ & Type & Warp \\
    &             & (1996) ID     &          &     &          &          & &\\
\hline                   
   15 &   733.9,271.3 & 3-923.0 & 24.70  &  &  0.40  &   0.49 &   Scd & 6 \\
   16 &  1342.8,274.7 & 3-925.0 & 24.93  &  &  0.92  &   0.88 &   Irr & U \\
   55 &  3893.6,380.4 & 4-976.2 & 23.32  &  &  0.20  &   0.15 &   Irr & \\
   81 &  3232.6,433.0 & 4-618.0 & 23.94  &  &  0.64  &   0.60 &   Irr & \\
   112&  2764.7,508.1 & 4-386.0 & 26.09  &  &  1.08  &   0.87 &   Irr & \\
   120&  1368.7,534.7 & 3-786.0 & 24.27  &  &  1.60  &   1.60 &   Irr & \\
   153&   673.6,598.0 & 3-761.1 & 24.94  &  &  0.76  &   0.69 &   Irr & \\
   246&  2154.6,809.8 & 4-70.0  & 26.99  &  &  1.48  &   1.69 &   Irr & \\
   273&  2107.1,872.3 & 4-33.0  & 22.47  & 0.905&0.68&   0.80 &   Ell & \\
   304&  1747.1,951.1 & 3-593.1 & 25.60  &  &  1.760  &   1.79 &   Scd&  U\\
   319&  2178.7,989.2 & 4-85.1  & 24.41  & 0.961&0.92&   0.90 &   Irr & U\\
   450&   929.2,1329.3 & 3-398.1 & 24.25  &  &  0.72  &   0.67 &   Irr& \\
   476&  2430.9,1390.5 & 4-232.12& 22.77  & 0.421&0.48&   0.45 &   Irr& \\
   506&  1899.5,1472.4 & 3-331.0 & 24.47  & 0.751&1.08&   0.75 &   Irr& \\
   534&  3812.9,1545.4 & 4-926.1 & 23.82  &  & 1.04   &   0.85 &   Sbc& 5\\
   632&  3483.8,1771.9 & 4-745.0 & 25.80  &  & 1.60   &   1.88 &   Irr& \\
   671&  2166.4,1877.4 & 4-89.0  & 23.93  & 0.681&0.76&   0.65 &   Irr& \\
   716&  3849.0,2034.0 & 4-946.0 & 22.61  & 0.944&0.68&   0.45 &   Sbc& 5\\
   727&  3792.2,2078.6 & 4-916.0 & 23.93  & 0.904&0.16&   1.88 &   Scd& 6\\
   733&  3628.1,2104.3 & 4-829.1 & 25.04  &  &   0.96 &   0.91 &   Irr& \\
   749&  1458.6,2208.5 & 2-339.1 & 22.60  & 0.851&0.88&   0.95 &   Sbc& U\\
   774&  2521.0,2361.5 & 1-34.1  & 21.22  & 0.485&0.68&   0.40 &   Sbc& \\
   805&   172.2,2492.4 & 2-1022.0& 25.33  &  &    0.72&   0.59 &   Irr& \\
   817&  2793.4,2550.5 &1-57.11111&22.19  & 0.485&0.68&   0.50 &   Sbc& 4\\
   886&  1835.2,2838.4 & 2-133.0 & 25.25  &  &    1.00&   0.94 &   Irr& \\
   888&   196.0,2852.5 & 2-1018.0& 23.70  & 0.559&0.64&   0.56 &   Irr& \\
   898&  1958.2,2909.9 & 2-65.1  & 25.89  &  &    0.92&   0.95 &   Irr& 6\\
   899&   181.2,2923.1 & 2-1023.1& 22.94  & 0.564&0.68&   0.60 &   Scd& \\
   938&   765.0,3128.9 & 2-702.1 & 23.03  & 0.557&0.52&   0.55 &   Scd& \\
   979&   327.1,3397.0 & 2-950.0 & 23.45  & 0.517&0.44&   0.50 &   Scd& 4\\
  1027&  1230.6,3650.7 & 2-432.0 & 23.88  &  &   0.92 &   0.87 &   Irr& \\
  1031&  1590.7,3680.9 & 2-270.2 & 25.08  &  &   1.20 &   0.93 &   Scd& \\
\hline
\end{tabular}
\end{center}
\end{table*}

\begin{table*}
\caption{Edge-on galaxies in the HDF-S}
\begin{center}
\begin{tabular}{|c|c|c|c|c|c|}
\hline
Name & $I_{814}$ & $z_{fsoto}$ & $z_{font}$ & Type & Warp \\
\hline                   
 SB-WF-0501-0760 &    25.07  &          1.11&    1.31&   Irr & \\
 SB-WF-0564-0515 &    25.27  &          1.01&    1.03&   Irr & \\
 SB-WF-0566-1028 &    25.67  &          1.79&    2.52&   Scd & \\
 SB-WF-0578-2216 &    23.16  &          1.08&    0.88&   Scd & 6\\
 SB-WF-0747-1929 &    25.31  &          1.07&    0.88&   Irr & \\
 SB-WF-0932-2115 &    25.58  &          1.03&    0.88&   Scd & \\
 SB-WF-1085-0562 &    25.51  &          1.07&    1.19&   Irr & \\
 SB-WF-1404-2925 &    23.09  &          0.54&    0.45&   Sbc & \\
 SB-WF-2340-1898 &    24.21  &          0.55&    0.56&   Irr & \\
 SB-WF-2353-1914 &    24.36  &          0.30&    0.31&   Scd & \\
 SB-WF-2504-3117 &    24.80  &          1.20&    1.09&   Irr & 4\\
 SB-WF-2661-4090 &    24.10  &          0.92&    0.92&   Irr & \\
 SB-WF-2691-2966 &    25.05  &          0.49&    0.42&   Irr & \\
 SB-WF-2816-2895 &    26.04  &          0.47&    0.63&   SB1 & \\
 SB-WF-3053-1963 &    26.07  &          1.18&    0.91&   Scd & 5\\
 SB-WF-3329-2173 &    24.54  &          0.27&    0.32&   Scd & \\
 SB-WF-3458-1026 &    23.40  &          0.69&    0.69&   Scd & 6\\
 SB-WF-3685-2040 &    26.63  &          2.43&    2.03&   SB1 & \\
\hline
\end{tabular}
\end{center}
\end{table*}

\subsection{Completeness}

Visual classification is possible for relatively large galaxies
only. Thus one can expect that our sample is a diameter-limited
sample. In order to estimate the completeness limit we applied
the well-known V/V$_{max}$ test (Schmidt 1968). For a diameter-limited 
sample the ratio between the volume of the sphere determined
by the object distance (V) and the maximum volume of the sphere
in which the object could lie and still have been included in the
sample (V$_{max}$) is V/V$_{max}$=(d$_0$/d)$^3$, where d 
is the apparent angular diameter of the galaxy and d$_0$ is the
minimum diameter of the sample selection criterion (e.g. Thuan \&
Seitzer 1979). For objects distributed uniformly in space the
average value of V/V$_{max}$ should be 0.5$ \pm 1/\sqrt{12 \times N}$,
where $N$ is the number of sample galaxies.

We determined the apparent major axis angular diameter (d) within 
the isophote $\mu(I_{814})=26.0$ for all the sample galaxies. Then, 
changing d$_0$  and computing the V/V$_{max}$ ratio, we found that 
for d$_0$=1.$''$33 $\langle$V/V$_{max}$$\rangle$=0.50$\pm$0.04. 
Therefore, our sample is nearly complete for the galaxies with diameter
d$>$1.$''$3. (Note that we used the isophotal sizes of galaxies for the 
V/V$_{max}$ test. This is not quite correct due to the influence of the
cosmological dimming and $K$--correction on the apparent sizes of
galaxies. Note also that surface brightness evolution 
-- e.g. Lilly et al. 1998 -- must partially compensate
the above mentioned effects.)

Only five galaxies (n112, n246 in the HDF-N and s0932, s2504, s2816
in the HDF-S) show major axis diameters smaller than 1.$''$3. We will
consider the rest of the sample -- 45 galaxies -- as a complete
sample.

\subsection{General characteristics}

Fig.1 shows the distribution of the selected objects within the
HDF-N and HDF-S. Uniform angular distribution of galaxies is evident in
the figure. HDF-N contains more edge-on galaxies (32) than HDF-S
(18). This difference probably reflects the general excess of galaxies
in the northern field at $z \leq 1$ (Gwyn 1999).

\begin{figure*}
\centerline{\psfig{file=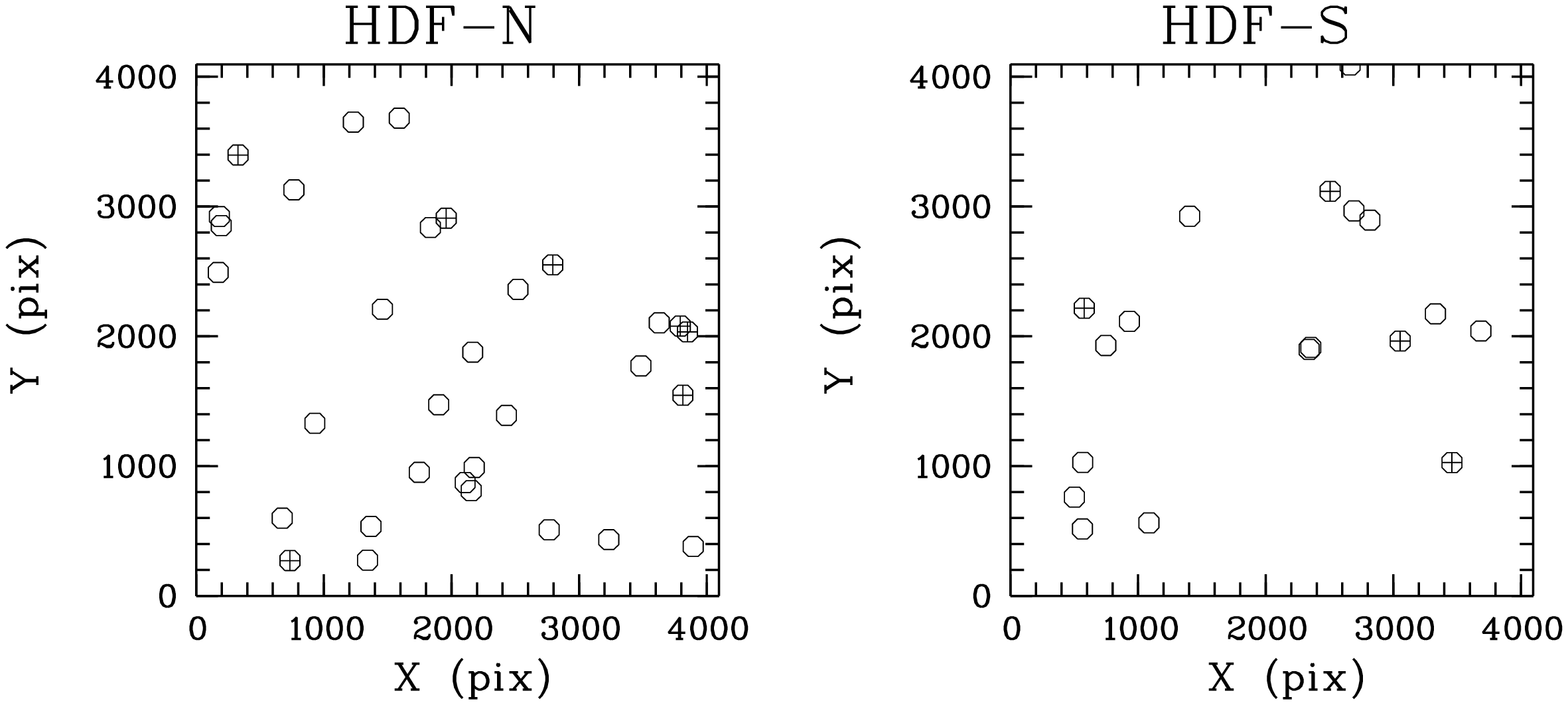,width=14.0cm,angle=-90,clip=}}
\caption{Distribution of edge-on galaxies within the two fields. 
Crossed circles -- galaxies with possible warps (see tables 1 and 2).}
\end{figure*}

We present available spectroscopic and photometric redshifts of
the galaxies in tables 1 and 2. Unfortunately, we have spectroscopic
redshifts for only about half of the objects (14) in the HDF-N. For
the remaining galaxies we will use the mean photometric redshifts from the 
catalogues of Fern\'{a}ndez-Soto et al. (1999) and Fontana et al. (2000). 
The edge-on orientation and, therefore, strong internal absorption
can distort the observed spectral energy distribution of a galaxy and
can introduce some systematic effects in the derived photometric
redshift. To check for this possible effect, we compared two catalogues of 
photometric redshifts with spectroscopic data.
The two sets of photometric redshifts demonstrate relatively good
agreement (Fig.2): \\
$\langle z_{fsoto}-z_{font} \rangle = +0.00 \pm 0.31(\sigma)$ ($N$=50) or \\
$\langle z_{fsoto}-z_{font} \rangle = +0.05 \pm 0.13$ ($z \leq 1.8$, $N$=46). \\
The comparison of spectroscopic and photometric redshifts does not
show any systematic effects (Fig.2): \\
$\langle z-z_{fsoto} \rangle = +0.02 \pm 0.26$ ($N$=14), \\
$\langle z-z_{font} \rangle = -0.03 \pm 0.31$ ($N$=14). \\
The largest differences between photometric and spectroscopic $z$
are for n716 and n727. These two galaxies are very close in 
projection (2.$''$9 or 17 kpc at $z=0.92$) and show close spectroscopic 
$z$ (see table 1). We suppose that these galaxies form a real physical
pair and that their spectral energy distributions are strongly
distorted by tidally-induced star formation.

\begin{figure*}
\centerline{\psfig{file=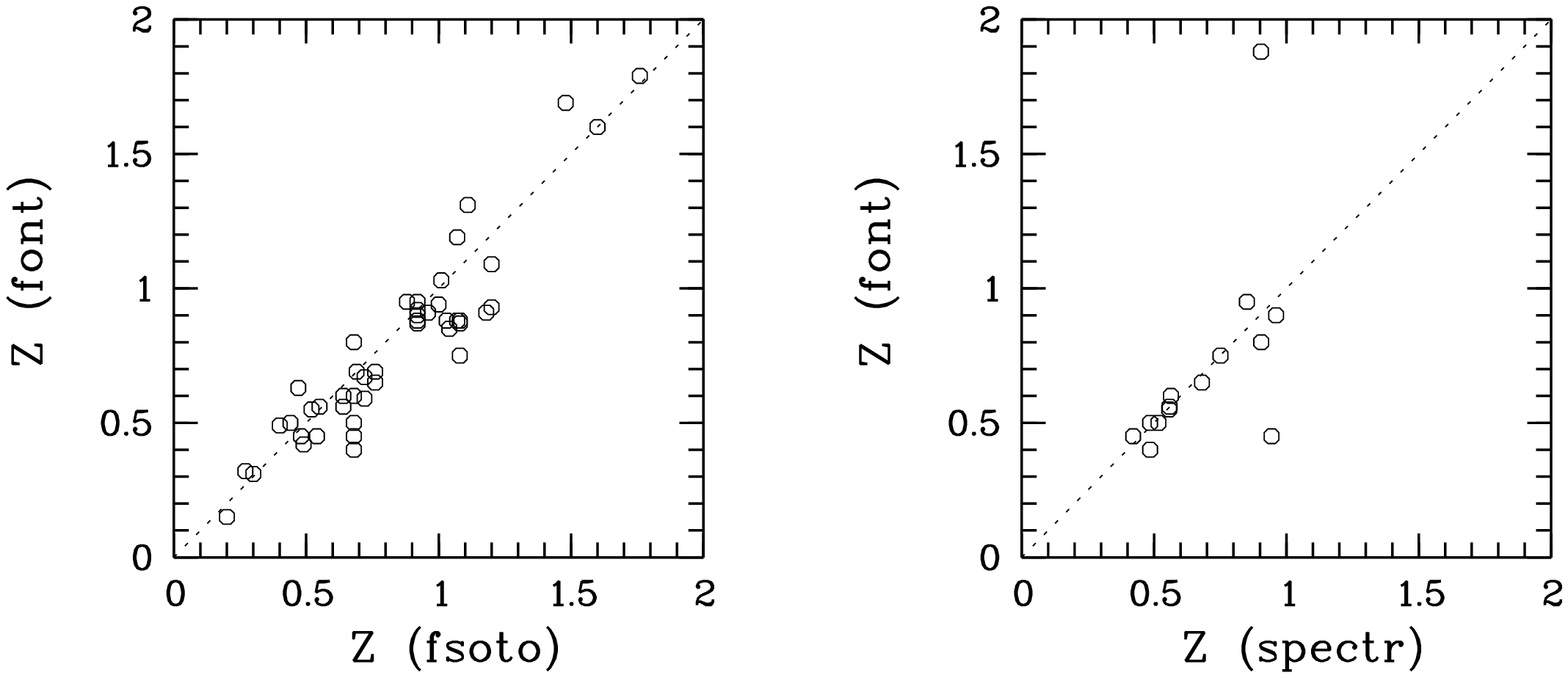,width=15.0cm,angle=-90,clip=}}
\caption{Left: comparison of photometric redshifts of our sample
edge-on galaxies according to Fern\'{a}ndez-Soto et al. (1999) and
Fontana et al. (2000) catalogues. Right: comparison of spectroscopic
and photometric redshifts for 14 edge-on galaxies.}
\end{figure*}

Fig.3 presents the redshift distribution of edge-on galaxies. 
The distribution is peaked at $z \sim 0.8-0.9$ (table 3) 
and is located in the
low-redshift wing of a global redshift distribution of galaxies
in the deep fields (bottom part of Fig.3). This is probably 
explained by our diameter-limited selection of the sample.

\begin{figure}
\centerline{\psfig{file=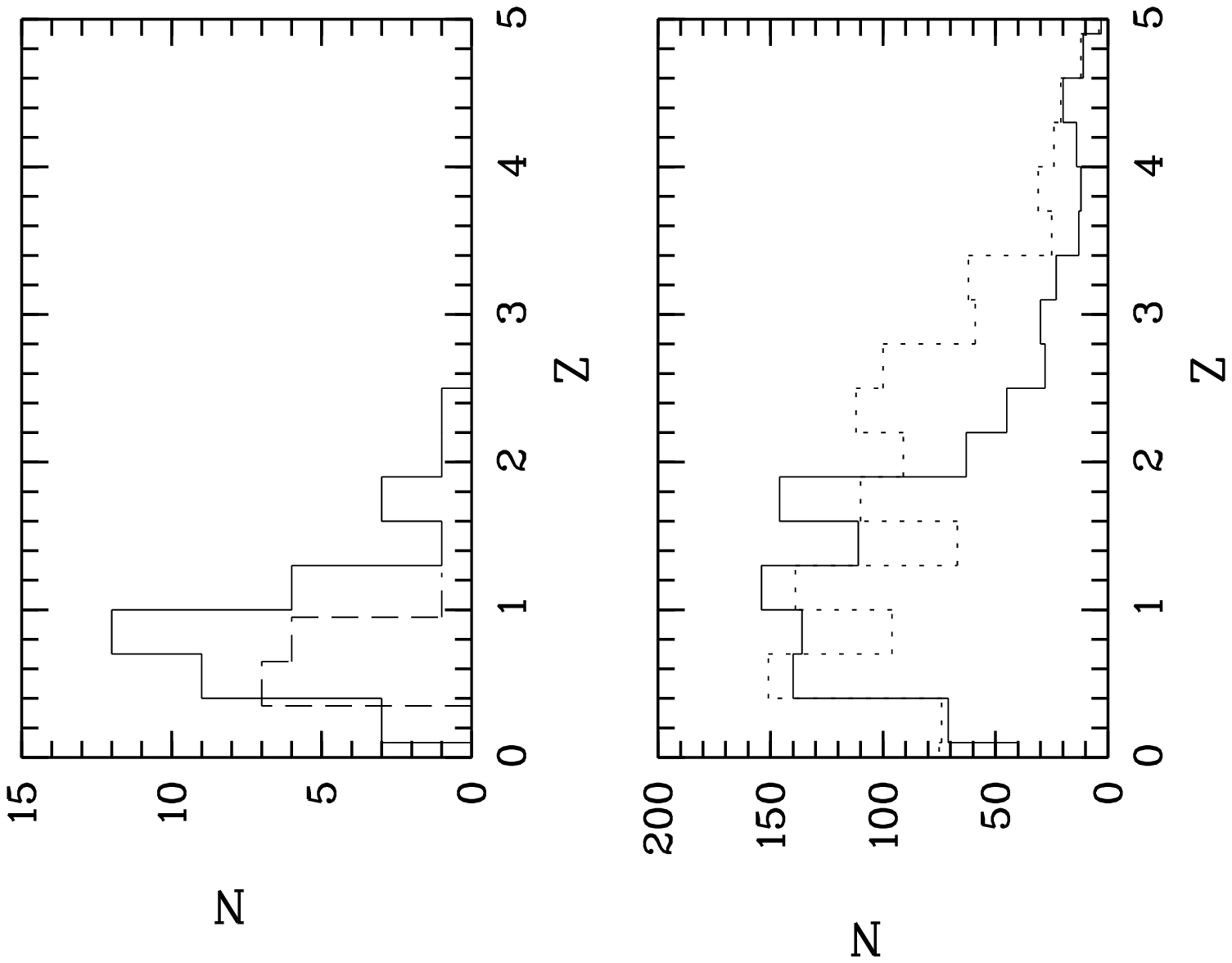,width=9.0cm,angle=-90,clip=}}
\caption{Top: redshift distribution of edge-on galaxies
(solid line shows galaxies with photometric redshifts, dashed line
-- with spectroscopic determinations). Bottom: distributions of
the photometric redshifts in the HDF-N (solid line) and
HDF-S (dashed line) according to Fern\'{a}ndez-Soto et al. (1999) and
Fontana et al. (2000) catalogues.}
\end{figure}

Fig.4 shows the distribution of apparent axial ratios -- $b/a$ -- 
of the sample objects. The $b/a$ values were estimated from the
ellipse-fitting of outer contours of the galaxies. 
Our $b/a$ estimates are very close to those published in 
the Williams et al. (1996) catalogue. The mean ratio 
$\frac{b/a({\rm pres. work})}{b/a({\rm Williams})}$ is 
1.01$\pm$0.21$(\sigma)$. The average value of axial ratio
-- 0.32 (table 3) -- significantly exceeds the $b/a$ ratio
for local spiral galaxies. According to Guthrie (1992) 
the true axial ratio of spiral disks is only 0.11--0.13.
Assuming a very conservative true axial ratio of our sample of edge-on
galaxies $b/a=0.2$, we can estimate that the observational value (0.32)
corresponds to a rather large inclination of the galaxies 
($i \sim 75^{\rm o}$). Such a large departure from edge-on orientation
must lead to very strong asymmetry of the surface brightness 
distributions along the minor axis. However we do not see any such significant
asymmetries among the sample galaxies. Another possible
reason is that the sample objects possess a large intrinsic
axial ratio ($b/a \approx 0.3$) -- about twice that of
local spirals. This conclusion is consistent with the facts that
1) interaction and merger rates were significantly higher in the
past (see Le F\'{e}vre et al. 2000, Reshetnikov 2000 and references
therein) and 2) galaxy interactions lead to thickening
of the stellar disks (Reshetnikov \& Combes 1997; 
Schwarzkopf \& Dettmar 2000; see also Karachentsev \& Karachentseva 1974).
 Of course, our conclusion is extremely tentative since we
must check possible influence of the PSF on the observed thicknesses
of edge-on galaxies in the HDF.

\begin{table}
\caption{Mean characteristics of the sample}
\begin{center}
\begin{tabular}{|c|c|c|c|}
\hline
Parameter & HDF-N & HDF-S & N+S  \\
\hline                   
Number    & 32    &  18   & 50 \\
$\langle I_{814} \rangle$ & 24.15$\pm$1.30($\sigma$) &
24.88$\pm$1.02 & 24.41$\pm$1.24 \\
$\langle z \rangle$ & 0.85$\pm$0.38 & 0.95$\pm$0.54 & 0.89$\pm$0.44 \\
$\langle b/a \rangle$ & 0.31$\pm$0.07 & 0.34$\pm$0.08 & 0.32$\pm$0.08 \\
$\langle M_B \rangle$ & --18.1$\pm$1.2 & --17.7$\pm$1.4 & 
--17.9$\pm$1.3\\
\hline
\end{tabular}
\end{center}
\end{table}

\begin{figure}
\centerline{\psfig{file=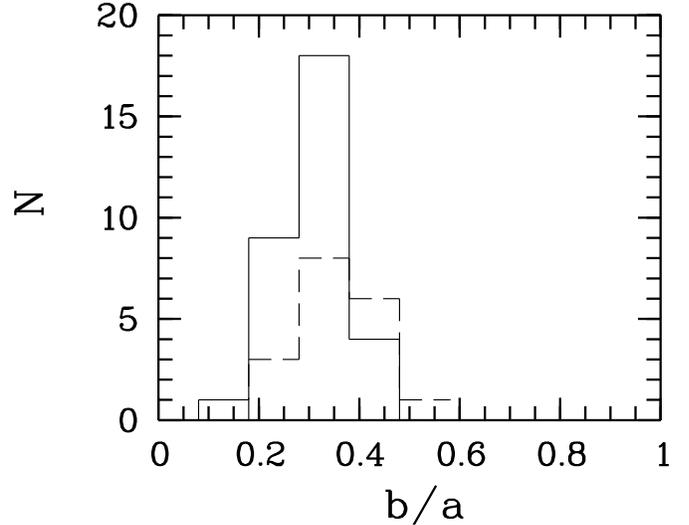,width=9.0cm,angle=-90,clip=}}
\caption{Apparent axial ratios: HDF-N (solid line) and
HDF-S (dashed line).}
\end{figure}

The mean {\it isophotal} diameter of the sample galaxies within
$\mu(I_{814})=26.0$ is 2.$''$15$\pm$0.$''$9 or 12.9 kpc at $z=0.9$.

The rest-frame absolute magnitudes in the $B$ passband were
estimated following Lilly et al. (1995): \\
$M(B) = I_{814} - 5\,{\rm log}(D_L/10\,{\rm pc}) + 2.5\,{\rm log}(1+z) + 
(B-I_{814,z})$,
where $D_L$ is the luminosity distance and $B-I_{814,z}$ is the $K$-correction
color. For our purposes we used the $B-I_{814,z}$ term for Irr galaxy
(Lilly et al. 1995). At high redshifts, this term is small since
$I_{814}$ is redshifted down toward the rest-frame $B$ filter
(they coincide at $z=0.83$). The resulting mean rest-frame absolute
magnitude is $M(B) \approx -18$ (table 3). After correction for internal 
absorption the absolute magnitude must be $\approx -19$. Therefore,
our sample objects are sub-$L^*$ galaxies. 

The mean parameters of the sample are summarized in table 3.

\section{Statistics of warped galaxies}

\subsection{Could we recognize local warps at high redshifts?}

Searching for optical warps in the disks of distant galaxies
is a difficult enterprise.
In local galaxies optical warps become detectable in the 
outer parts of their disks. For instance, from a sample of
45 spirals with measureable warps, Reshetnikov (1995) estimated
that the average value of the surface brightness level 
at which optical warping becomes detectable is $\mu(R)=23.0 \pm 0.8$ 
(in the Cousins $R$ passband) or $\mu(B) \approx 24$. 
(After correction for the line-of-sight
integration, this level corresponds to $\mu(B) \approx 25$ for face-on galaxy.) 
Moving such a typical object at $z=0.9$ (average redshift of our sample),
this level will fall to $\mu(I_{814}) \approx 26$ (see Lilly et al. 1995
and note that the $K$-correction term is close to zero at this redshift).
This surface brightness corresponds to $\sim 1.5\sigma$ limiting
isophote of the deep fields and, therefore, we cannot recognize
such a typical optical warp at $z \sim 1$.

In order to directly illustrate this conclusion we have 
redshifted the $B$ band image of the local Sb spiral ESO~235-G53,
which has a remarkable optical warp. The image of the galaxy was taken by
R. de Grijs with the Danish 1.54-m telescope at La Silla, Chile,
in July 1994. The prominent warp starts at
the radius $\approx (1.2-1.7)h$ 
and reaches $\approx 10^{\rm o}$ near the end of the disk
(de Grijs 1997). Using the method described in Giavalisco et al. (1996),
we placed ESO~235-G53 at $z=0.83$ where redshifted $B$ corresponds
to $I_{814}$. The result of the redshifting is shown in Fig.5. 
(Note that this is just an illustrative figure and we faded the whole image
including the foreground stars.)
The result demonstrates that even such a strong warp cannot be 
detected.

\begin{figure}
\centerline{\psfig{file=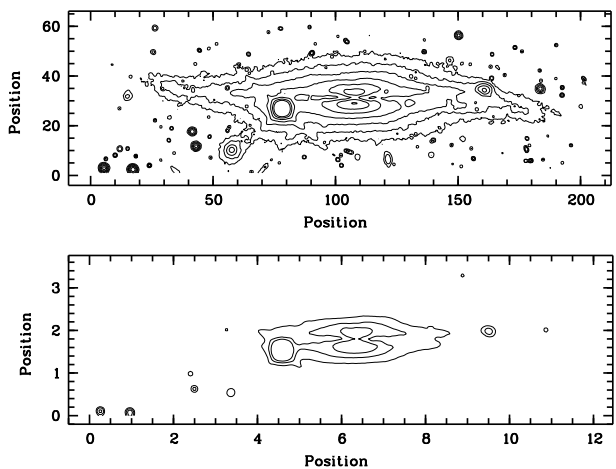,width=8.7cm,angle=-90,clip=}}
\caption{Top: contour map of ESO~235-G53 in the $B$ passband.
The faintest contour corresponds to $\mu(B)=27.0$. Bottom: 
contour map of the same galaxy moved at $z=0.83$. 
The faintest contour corresponds
to $\mu(I_{814})=26.5$ or $\sim1\sigma$ limiting
isophote of the HDF. Isophote step is 0.$^m$75. Both axes are
in arcseconds.}
\end{figure}

We can conclude from the above exercise that among distant
galaxies we can recognize only extremely strong and bright optical warps. 
Any statistics of warps in the deep fields will give a strict lower
limit of the true fraction due to a huge observational bias
against finding them.

\subsection{Selection of warped galaxies}

The original $I_{814}$-band images and contour maps of all the
sample galaxies were examined independently by three of 
us (E.B., F.C. and V.R.). On the basis of eyeball classification,
they assigned an "index of warp" (IW) for each object:
0 -- no warp, 2 -- a certain integral-shaped warp, and 1 --
intermediate case. We also marked U-shaped warps,
when the plane bends towards the same direction on both sides 
(Reshetnikov \& Combes 1998) like the letter "U". Then, we summed
the personal indices for each object and decided to consider 
those galaxies with total IW=6 (index 2 for each observer)
as true warps and IW=4-5 as good candidates for warped disks. 
There are 5 galaxies with true warps, and 6 good candidates
in our sample of 50. Our classification for the galaxies with IW$\geq$4
is presented in the last columns of tables 1 and 2. Contour
maps of all galaxies with certain or possible warps
are displayed in Fig.6 and 7.
 Note that these are just illustrative contour maps and they 
do not give the best  impression of possible warps. A detailed
inspection must be made on the original pictures.

These 11 galaxies were processed with a special software, written to
automatically detect and measure some quantitative characteristics
of warps. The software works essentially as the WIG program described by 
Jim\'enez-Vicente et al. (1997). First the images are rotated so the
galaxies are roughly in a horizontal position and the pixels with a
S/N$>$3 are selected. Then the warp curve
(mean deviation from the equatorial plane for each radius) is 
calculated by fitting a gaussian to each column. Outliers are 
removed using S/N and continuity criteria. The flat part of the
galaxy is defined by the average position angle of the inner FWHM of
the galaxy disk.

The software detects a warp if there are more
than five consecutive pixels with deviations larger than twice the
standard deviation in the flat part. In this case the first of those
pixels is taken as the starting radius of the warp. Then, the warp
angle is determined for both sides (defined as the angle between the 
equatorial plane and the
line joining the center with the point with the largest deviation from
the equatorial plane in the warp curve) and averaged to give the
average warp angle.  

As a result, we confirmed that 9 of 11 galaxies demonstrate
integral-like shapes according to our criteria.

Table 4 summarizes the general characteristics of
optical warps in the sample according to the software.
The columns are as follows: $\psi$ -- mean warp angle,
R$_{\rm warp}$ -- starting point of the warp, and 
R$_{\rm warp}$/$h$ -- starting
point of the warp to exponential scalelength ratio
(uncertain values are marked by colons).

\begin{table}
\caption{Characteristics of optical warps}
\begin{center}
\begin{tabular}{|c|c|c|c|}
\hline
Galaxy & $\psi$  & R$_{\rm warp}$ & R$_{\rm warp}$/$h$ \\
       & $(^{\rm o})$&  (arcsec) & \\
\hline                   
n15    & 4.5        & 0.58  & 1.5 \\
n716   & 1.7        & 0.62  & 1.0:\\
n727   & 5.5        & 0.60  & 1.5 \\
n817   & 5.3        & 0.88  & 1.8 \\
n898   & 5.6        & 0.52  & 1.1:\\
n979   & 5.7        & 0.62  & 1.9 \\
s0578  & 11.4       & 1.08  & 1.6:\\
s3053  & 9.2        & 0.66  & 1.2:\\
s3458  & 4.5        & 0.38  & 1.0:\\
\hline
\end{tabular}
\end{center}
\end{table}

\subsection{Statistics of warps}

The measured mean warp angle is
$\langle \psi \rangle$=5.$^{\rm o}$9$\pm$2.$^{\rm o}$8($\sigma$) or,
excluding two extreme cases (n716 and s0578),  
5.$^{\rm o}$8$\pm$1.$^{\rm o}$6. This value is larger than reported
by Reshetnikov (1995) or Reshetnikov \& Combes (1998) -- 
3$^{\rm o}$--4$^{\rm o}$ -- for local spirals. The mean warp angle
for distant objects is also larger than for local galaxies with
strongest optical warps -- 4.$^{\rm o}$8$\pm$1.$^{\rm o}$3
(Reshetnikov \& Combes 1999).

The observed fraction of optical warps in the HDF-N and HDF-S is
9/50 = 0.18$\pm$0.06 (Poisson error). Considering the ''complete'' sample
of 45 galaxies, this fraction is 9/45 = 0.20$\pm$0.07. The observed
fraction is certainly an underestimate of the true fraction (see sect.3.1).

There is a possibility of confusing a warped disk with a slightly 
inclined (from edge-on) disk with a contrasted 2-spiral pattern
(see the simulations in Reshetnikov \& Combes 1998). This type of
confusion should be less important at high redshift, since
grand-design spirals appear to be rarer for  z$\geq$0.3 or even absent in the past
(cf van den Bergh et al. 2000).  Possible projection of
small background objects, for instance for n898 and s0578, can be rejected on
the basis of smoothness and symmetry of the warp curves. 
However it is possible that in both cases we observe accretion
of small companions by a large spiral galaxy.

To obtain a rough estimate of the true fraction of distant warps, 
let us assume that warped disks at $z \sim 1$ obey the
same distribution of warp angles as the $z=0$ spirals -- $\psi^{-5}$
(Reshetnikov \& Combes 1998, 1999). Excluding two extreme cases with
$\psi=1.^{\rm o}7$ and 11.$^{\rm o}$4, we have 7 galaxies with
$\psi=4.^{\rm o}5-9.^{\rm o}2$. Assuming that we selected all
spirals in the two deep fields with $\psi=4.^{\rm o}5-9.^{\rm o}2$, 
one can infer that the true number of galaxies with
$\psi \geq 2.^{\rm o}5$ must be 31 or $\sim$2/3 of all galaxies
(2.$^{\rm o}$0--2.$^{\rm o}$5 is the detection limit of optical
warps in our survey of local spirals -- Reshetnikov \& Combes 1998). 
Therefore, the actual fraction of warped disks at $z \sim 1$
is the same or even larger than at $z=0$. Given the
geometrical factor (warps along the line of sight are
not detected), practically all distant disks are warped.

To draw a clearer conclusion about possible warp evolution with
redshift, let us compare the frequencies of strongly warped disks
at $z=0$ and in the deep fields. In the two fields we found 8 galaxies
with $\psi \geq 4.^{\rm o}5$ (table 4). What is the equivalent fraction 
in the local universe? About 40\% of spiral galaxies 
at $z=0$ reveal S-shaped optical warps with $\psi \geq 2^{\rm o}$
(Reshetnikov \& Combes 1998). Adopting the $\psi^{-5}$ law,
one can estimate that the local frequency of spirals with
$\psi \geq 4.^{\rm o}5$ is 0.7\%. Then, using the method described
in Reshetnikov (2000), we found that the volume density of
strong warps changes $\propto (1+z)^m$ with $m=3.8$. Assuming
Poisson errors for galaxy numbers, we have $m=3.8^{+0.6}_{-1.0}$
(owing to the uncertainty in the local frequency of strong warps, the actual
error on $m$ may be larger.) Comparing frequencies of galaxies with
$\psi \geq 5.^{\rm o}0$, we obtain $m=4.0^{+0.8}_{-1.5}$.
Our data therefore demonstrate a fast evolution of the volume
frequency of {\it strong} optical warps at $z \sim 1$. 
The galaxies are therefore more frequently warped, and
distant warps are brighter and stronger. 
The rate of the evolution is
consistent with the increase of galaxy interactions and mergers
(Le F\'{e}vre et al. 2000, Reshetnikov 2000), which tends
to suggest that the source of warping is precisely
tidal interaction.
Schwarzkopf \& Dettmar (2001) estimate that 
tidal perturbations generate the strongest warps.

\begin{figure*}
\centerline{\psfig{file=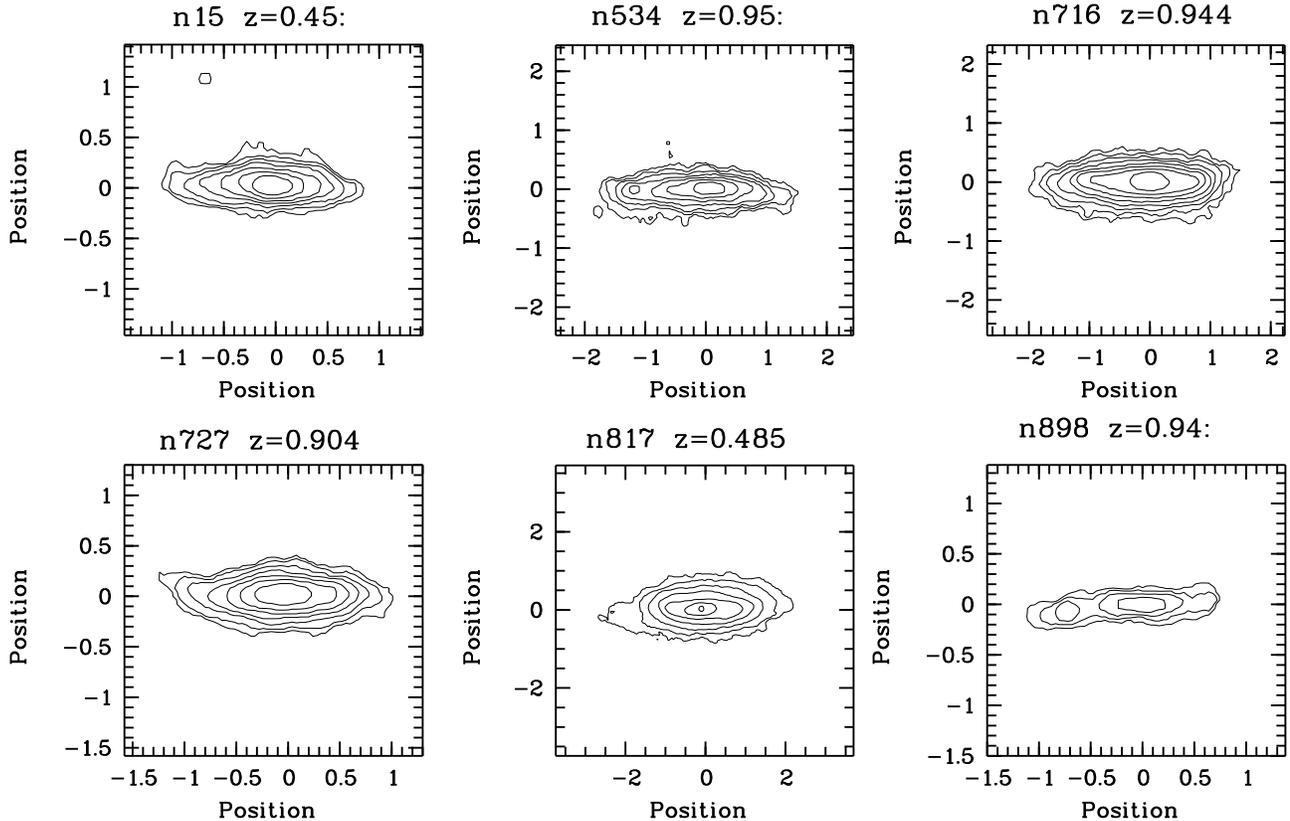,width=18.0cm,angle=-90,clip=}}
\caption{Contour plots of galaxies with certain or 
possible warps (see tables 1 and 2).
Both axes are in arcseconds (size of 1 pixel of the original fields
is 0.$''$04). The faintest contour corresponds to
$\mu(I_{814})=26.0$, isophotes step is 0.$^m$44 (0.$^m$75 for n817).
 Uncertain values of $z$
obtained as the mean of two photometric redshifts (see tables 1 and 2)
are marked as (:).}
\end{figure*}

\begin{figure*}
\centerline{\psfig{file=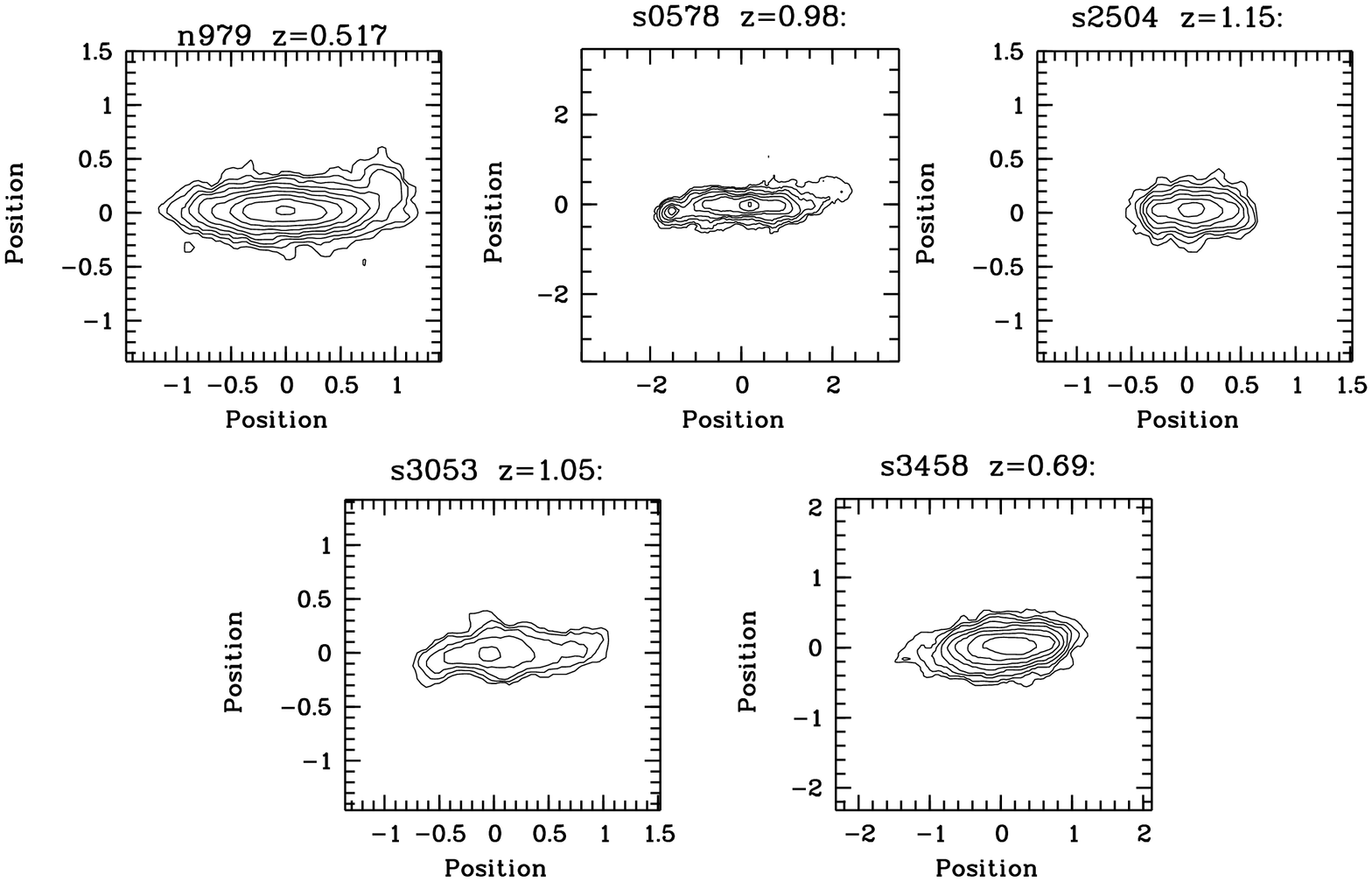,width=18.0cm,angle=-90,clip=}}
\caption{Contour plots of galaxies with certain or possible warps (continued).}
\end{figure*}

As one can find from Table 1, the mean projected starting point of
the stellar warps we detected is 
$\langle {\rm R_{\rm warp}}  \rangle$=3.7$\pm$1.3 kpc or, expressed 
in exponential scalelengths, 1.4$\pm$0.3\,$h$. 
According to de Grijs (1997), the same value for local spirals
is 2.1$\pm$1.0\,$h$.

No coherence in the orientation of warps was found. This is not
surprising, taking into account the large distances involved, of the
order of gigaparsecs. Galaxies with close redshifts were too few to
be of any consequence. This phenomenon
cannot be excluded, however (Battaner et al. 1991, Zurita \&
Battaner 1997).

\section{Discussion and conclusions}

The result of the present statistical study is that 
warps were very frequent in the past, and they were
of larger amplitude. One obvious interpretation comes from
galaxy interactions that were much more frequent, since they
represent an efficient trigger of warps.
Indeed the frequency of galaxy pairs and the frequency of mergers
is observed to increase as a high power of time, or in (1+z)$^m$,
with $m = 3-4$ (Le F\'{e}vre et al. 2000).  This is also observed 
in the fraction of perturbed and distorted galaxies, for instance in the
frequency of tidal tails (Reshetnikov, 2000). However, a much
lower rate of mergers or mass accretion was observed for bright
and massive galaxies by Carlberg et al. (2000). This might be
related to the fact that the spheroids or bulge-dominated galaxies
appear later in redshift, at the expense of disk-dominated and 
irregular galaxies (Kajisawa \& Yamada, 2001).
This is expected in a
hierarchical cosmological scenario, where galaxies are formed
by successive mergers. Galaxies are then more frequent in the past,
per comoving volume. Moreover, due to expansion, the density
of galaxies was larger (increasing as $(1+z)^3$ for a homogeneous universe).
 On the other hand, galaxies were smaller, and the warps appear
to begin at a smaller radius. At this distance from the center,
the differential precessing rates of orbits is shorter, and the
life-time of warps should be smaller (roughly proportional
to R$_{\rm warp}$, for flat rotation curves).

Another possible triggering of warps is the action of an intergalactic
magnetic fields.
In  a highly conductive homogeneous universe, magnetic fields
increase as $(z+1)^2$ (Battaner \& Florido 2001), which means a factor
of 4 at $z=1$. However, at $z=1$, with a highly inhomogeneous universe
the condition of frozen-in magnetic field lines should better describe
the evolution of magnetic fields. 
We adopt $B\rho^{-2/3}$=constant. Then, for a present-day cluster overdensity of
10, the  magnetic field strength should be 
$2^{2/3}=1.6$ larger than today. Magnetic fields larger by a factor 
1.6-4 also constitute a potential explanation of the increased warping
amplitude in the past.

The infall of dark and visible matter either on halos (Ostriker \&
Binney 1989) or on disks,
producing reorientation of the galactic angular momentum, should have
been greater in the past. Therefore the infall mechanism could also
be a dominant effect to explain the high rate of $z \sim 1$ warps.

 Finally, we would like to stress that our results are obtained
for very small regions of the celestial sphere. Moreover, the 
statistics are based on small numbers. Therefore, further
extended studies of distant edge-on galaxies are highly needed
to develop our preliminary results.

\acknowledgements{
This work was begun while V.R. was visiting Paris, thanks to a one 
month grant from Paris Observatory. We would like to thank R. de Grijs 
and A.Guijarro for providing the image of ESO~235-G53 in the $B$ 
passband.}


\begin{thebibliography}{}
\bibitem{} Battaner, E., \& Florido, E. 2001, Fundamentals of Cosmic
Physics, 21, 1
\bibitem{} Battaner, E., Florido, E., \& S\'{a}nchez-Saavedra, M.L. 1990,
A\&A, 236, 1 
\bibitem{} Battaner, E., Garrido, J.L., S\'{a}nchez-Saavedra, M.L., \&
Florido, E. 1991, A\&A, 251, 402
\bibitem{} Binney, J. 1992, ARAA 30, 51
\bibitem{} Binney, J., Jiang, I., \& Dutta, S. 1998, MNRAS, 297, 1237 
\bibitem{} Bosma, A. 1981, AJ, 86, 1825 
\bibitem{} Briggs, F. H. 1990, ApJ, 352, 15 
\bibitem{} Carlberg R.,  Cohen, J. G., Patton, D. R. et al. 2000, ApJ 532, L1
\bibitem{} Cohen, J.G., Hogg, D.W., Blandford, R., et al. 2000, ApJ, 538, 29
\bibitem{} Dav{\' e}, R., Hernquist, L., Katz, 
N., \& Weinberg, D. H. 1999, ApJ, 511, 521 
\bibitem{} de Grijs, R. 1997, Edge-on Disk Galaxies, Thesis, Groningen
\bibitem{} Fern\'{a}ndez-Soto, A., Lanzetta, K.M., \& Yahil, A. 1999,
ApJ, 513, 34
\bibitem{} Fontana, A., D'Odorico, S., Poli, F., et al. 2000, AJ, 120, 2206
\bibitem{} Giavalisco, M., Livio, M., Bohlin, R.C., et al. 1996, AJ, 112, 369
\bibitem{} Guthrie, B.N.G. 1992, A\&AS, 93, 255
\bibitem{} Gwyn, S.D.J. 1999, in Photometric Redshifts and High Redshift
Galaxies, Weymann, R., Storrie-Lombardi, L., Sawicki, M., \& Brunner, R.
eds., San Francisco, 61 [astro-ph/9907336]
\bibitem{} Hogg, D.W., Pahre, M.A., Adelberger, K.L., et al. 2000, ApJS, 127, 1
\bibitem{} Jiang, I. \& Binney, J. 1999, MNRAS, 303, L7 
\bibitem{} Jim\'{e}nez-Vicente, J., Porcel, C., S\'{a}nchez-Saavedra,
M.L., \& Battaner, E. 1997, A\&SS, 253, 225
\bibitem{} Fontana, A., D'Odorico, S., Poli, F., et al. 2000, AJ, 120, 2206
\bibitem{} Ferguson, H.C., Dickinson, M., \& Williams, R. 2000, ARAA, 38,667
\bibitem{} Karachentsev, I.D., \& Karachentseva, V.E. 1974, Astron. Zh., 51, 724
\bibitem{} Kajisawa M., Yamada T.: 2001, PASJ 53, 833
\bibitem{} Le F\'{e}vre, O., Abraham, R., Lilly, S.J., et al. 2000, MNRAS, 311, 565
\bibitem{} Lilly, S.J., Tresse, L., Hammer, F., et al. 1995, ApJ, 455, 108
\bibitem{} Lilly, S., Schade, D., Ellis, R., et al. 1998, ApJ, 500, 75
\bibitem{} Nelson, R. W. \& Tremaine, S. 1995, MNRAS, 275, 897 
\bibitem{} Oke, J.B. 1974, ApJS, 27, 21
\bibitem{} Ostriker, E.C., \& Binney, J.J. 1989, MNRAS, 237, 785
\bibitem{} Reshetnikov, V.P. 1995, A\&ATr, 8, 31
\bibitem{} Reshetnikov, V.P. 2000, A\&A, 353, 92
\bibitem{} Reshetnikov, V., \& Combes, F. 1997, A\&A, 324, 80
\bibitem{} Reshetnikov, V., \& Combes, F. 1998, A\&A, 337, 9
\bibitem{} Reshetnikov, V., \& Combes, F. 1999, A\&AS, 138, 101
\bibitem{} S\'{a}nchez-Saavedra, M. L., Battaner, E., \& Florido, E. 1990, MNRAS, 246, 458 
\bibitem{} Schmidt, M. 1968, ApJ, 151, 393
\bibitem{} Schwarzkopf, U., \& Dettmar, R.-J. 2000, A\&A, 361, 451
\bibitem{} Schwarzkopf, U., \& Dettmar, R.-J. 2001, A\&A, 373, 402
\bibitem{} Shang, Z. et al. 1998, ApJ, 504, L23 
\bibitem{} Thuan, T.H., \& Seitzer, P.O. 1979, ApJ, 231, 680
\bibitem{} van den Bergh, S., Cohen, J. G., Hogg, D. W., Blandford, R. 2000, AJ 120, 2190
\bibitem{} Williams, R.E., Blacker, B., Dickinson, M., et al. 1996, AJ, 112, 1335
\bibitem{} Zurita, A., Battaner, E., 1997, A\&A 322, 86
\end{thebibliography}
\end{document}